\pdfoutput=1
\documentclass[twocolumn,preprintnumbers,amsmath,amssymb,letter,nofootinbib]{revtex4}
\usepackage{graphicx}
\usepackage{dcolumn}
\usepackage{bm}
\renewcommand{\arraystretch}{1.25}

\begin{document}

\title{Quantum Walks with Gremlin\footnote{Rodriguez, M.A., Watkins, J.H., ``Quantum Walks with Gremlin," GraphDay `16, 1(1), pages 1--16, Austin Texas, January 2016.}}

\author{Marko A. Rodriguez}
\affiliation{Director of Engineering, DataStax Inc. \\ Project Management Committee, Apache TinkerPop}

 \author{Jennifer H. Watkins}
 \affiliation{Information Systems and Modeling, Los Alamos National Laboratory}

\date{\today}

\begin{abstract}
A quantum walk places a traverser into a superposition of both graph location and traversal ``spin." The walk is defined by an initial condition, an evolution determined by a unitary coin/shift-operator, and a measurement based on the sampling of the probability distribution generated from the quantum wavefunction. Simple quantum walks are studied analytically, but for large graph structures with complex topologies, numerical solutions are typically required. For the quantum theorist, the Gremlin graph traversal machine and language can be used for the numerical analysis of quantum walks on such structures. Additionally, for the graph theorist, the adoption of quantum walk principles can transform what are currently side-effect laden traversals into pure, stateless functional flows. This is true even when the constraints of quantum mechanics are not fully respected (e.g.~reversible and unitary evolution). In sum, Gremlin allows both types of theorist to leverage each other's constructs for the advancement of their respective disciplines.
\end{abstract}

\maketitle

\raggedbottom

\section{Introduction}

Gremlin is a graph traversal machine and language developed and distributed by the Apache TinkerPop project of the Apache Software Foundation \cite{gremlinmachine:rodriguez2015}.\footnote{Apache TinkerPop available at \texttt{http://tinkerpop.apache.org/}.} The Gremlin language is a human readable/writable graph programming language used to create Gremlin traversals (programs). Gremlin traversals are evaluated by the Gremlin traversal machine. The Gremlin traversal machine is a distributed virtual machine that can execute traversals over graphs contained on a single computer or represented across a multi-machine compute cluster. The traversal machine is agnostic to the underlying graph computing system and is supported by numerous OLTP/transactional graph databases and OLAP/batch graph processors.\footnote{The Gremlin traversal machine and language are analogous in many ways to the Java virtual machine and language \cite{jvm:lindholm1999}. For instance, both maintain a language and a machine, both are agnostic to the underlying evaluator (graph and operating system, respectively), and both allow other languages to compile to their respective machines' instruction sets.} Gremlin is used to create and analyze directed, binary, attributed, multi-graphs known as \textit{property graphs}. The expressivity of Gremlin, along with its Turing Complete \cite{compute:turing1937} nature, enables it to simulate discrete quantum walks. Unlike classical walks, a quantum walk places a traverser into a superposition of both locations and ``spins" across the graph \cite{aharonov:quantum1993,quantumintro:kempe2003}. Quantum walk theory provides a ``coin/shift"-model capable of universal quantum computing and as such, Gremlin may prove useful as a general-purpose quantum programming language when real-world quantum computers come to fruition \cite{quantlang:gay2006}.

Quantum walks on graphs have been studied extensively on lattices (one- and two-dimensional) and arguably less so on arbitrary undirected and directed graphs \cite{quantumdirected:montanaro2007}. In quantum walk theory, the traverser's location in the graph and its spin are represented in a geometric, complex vector space known as a Hilbert space. Due to quantum superposition, a traverser may be in multiple locations in the graph at the same time as well as have multiple spins at all such locations. When a traverser realizes multiple choices (e.g.~multiple incident edges to its current vertex location), the traverser is cloned/split and its respective spin components are projected accordingly. When multiple traversers merge (e.g.~multiple paths incoming to a particular vertex), all co-located traverser spins are summed via complex vector addition. Traverser spin merging can effect constructive (non-orthogonal) and destructive (orthogonal) wave interference because complex numbers are a superset of the reals ($\mathbb{R} \subset \mathbb{C}$) and can ``rotate" around the two-dimensional Argand plane \cite{spin:nachtergaele2004}. The potential for destructive interference makes the long run behavior of a quantum walk significantly different than its classical walk counterpart. In a classical walk, only constructive interference exists (known in Gremlin as ``bulking"). Moreover, it is the complexity of these wave interactions that make analytical solutions to quantum walks difficult beyond a two-dimensional lattice. Numerical simulation is currently the only feasible means to study quantum walks on complex graph structures of an arbitrary size.

This article is intended to be studied by both quantum and graph theorists and practitioners. In order to bridge these two domains, both disciplines are provided an introductory review which unifies the notational conventions of each. Quantum researchers should note the natural way in which the Gremlin language and machine can be used to represent and execute quantum walks on graphs at any scale.\footnote{The largest publicly known Gremlin processed graph in existence is the Amazon.com order fulfillment network which is approximately 1 trillion edges at the time of this writing.} Graph theorists should note how quantum traverser spin and interference semantics can be leveraged when designing graph algorithms/queries. Finally, the authors note that by studying discrete quantum walks, various insights were gleaned which may be applied to future versions of the Gremlin machine architecture. Section \S{2} will introduce quantum walk theory using standard Gremlin constructs. Section \S{3} will demonstrate one-dimensional quantum walks as well as the famous two-dimensional double-split screen experiment using Gremlin. Finally, section \S{4} will present a collection of property graph traversal motifs that leverage quantum walk concepts and which may prove fruitful in the advancement of Gremlin.

\section{Introduction to Quantum Walks}

A quantum graph walk can be understood as a propagation of an undulating wave across the set of elements of a graph. Given that waves can have crests and troughs, quantum walks yield constructive and destructive interference patterns analogous to those found in natural systems such as sound and water waves. Wave dynamics are not leveraged in classical walks and thus, quantum walks differ significantly in both their representation and semantics. The quantum walk machinery formalized herein not only enables the simulation of real-world quantum systems, but it also provides a new degree of freedom called ``spin" in the definition of a Gremlin traversal. This section will review the Gremlin graph traversal machine and language and discuss how \textit{quantum traversals} can be expressed. For the sake of discussion and notation simplicity, all examples will only use vertex locations. That is, the examples will never assume the traverser is located at an edge or property in the property graph. However, in practice and as allowed by Gremlin, traversers can exist at any such element in the property graph. Many of the introductory concepts around complex numbers, vector spaces, Dirac notation, and quantum computing in general is reviewed in lucid detail by \cite{quantumcomputing:yanofsky2008}.

\subsection{The Gremlin Traversal Machine and Language}

The Gremlin traversal machine is defined by three structures: a property graph $G$, a set of traversers $T$, and a traversal $\Psi$ \cite{gremlinmachine:rodriguez2015}. A property graph is defined as $G = (V, E \subseteq(V \times V), \lambda: (V \cup E) \times \Sigma^* \rightarrow U \setminus (V \cup E))$, where $V$ is the set of vertices, $E$ is the set of directed binary edges, and $\lambda$ is the property function that maps an element and character string ``key" to a non-relational object in the universal set $U$ (minus vertices and edges).\footnote{Properties can not reference vertices or edges and are typically used to reference primitives such as integers, doubles, strings, etc. Also note that $U$ will later be used to denote a unitary operator. The universal set $U$ is only used in this subsection.}  For example, the name property value of vertex $v$ is ``marko" and is denoted $\lambda(v,\text{name}) \mapsto \text{marko}$. The traversers $T$ execute the instructions specified in $\Psi$ in order to effect an algorithmic walk over $G$. The result of the computation is the location of all halted traversers in $T$ and any side-effect data structures yielded during the process (e.g. a \texttt{groupCount()}, \texttt{sum()}, \texttt{mean()}, etc. types of reduction). Every traverser in $T$ is composed of 6 properties which are discussed in detail in \cite{gremlinmachine:rodriguez2015}. For the presentation herein, only 4 traverser properties need to be reviewed.
\begin{enumerate}
 \item $\mu: T \rightarrow U$: the graph location of the traverser (e.g. a vertex, edge, property, etc.).
 \item $\psi: T \rightarrow \Psi$: the traversal location of the traverser (i.e. program counter).
 \item $\beta: T \rightarrow \mathbb{N}^+$: the ``bulk" of the traverser (i.e. its representative count).
 \item $\varsigma: T \rightarrow U$: the current ``sack" value of the traverser (i.e. a local mutable data structure). 
\end{enumerate}
Visually, a traverser $t \in T$ is a ``bundle" of local variables with a projection to a location in the graph $G$ and a projection to a location in the traversal $\Psi$ and is analogous, in many ways, to the definition of a central processing unit (CPU).
\begin{equation*}
G \longleftarrow\text{}_\mu \;\; \frac{t \in T}{\beta, \varsigma} \;\; \text{}_\psi\longrightarrow \Psi
\end{equation*}

The Gremlin language is used by humans to create traversals. Traversals are composed of primitive functions called steps. The instruction set of the Gremlin traversal machine is called the step library. Every step is either a \texttt{map}-, \texttt{flatMap}-, \texttt{filter}-, \texttt{sideEffect}-, or \texttt{branch}-step. For a review of these functional programming constructs, please see \cite{functional:hudak1989}. The Gremlin language supports three step composition motifs: $f \circ g \circ h \circ k$ (linear), $f(g \circ h) \circ k$ (nested), and $f(g,h) \circ k$ (parallel). The general idea is that traversers are generated at the start/leftmost step of the traversal and propagate from left to right while being modulated by each step along the way. Steps can grow (\texttt{flatMap}) or shrink (\texttt{filter}) the stream. The result(s) of the traversal are found at the end of the last/rightmost step. A collection of self-explanatory traversals are presented below to give a flavor of the Gremlin language. Later, when presenting quantum traversals, each traversal will be described in detail.

\begin{small}
\begin{verbatim}
// what has marko authored?
g.V().has(`name',`marko').out(`wrote')

// how many articles did marko write?
g.V().has(`name',`marko').out(`wrote').count()

// who are marko's coauthors?
g.V().has(`name',`marko').out(`wrote').in(`wrote')
 
// who are marko's coauthors that are not himself?
g.V().has(`name',`marko').as(`a').out(`wrote').
 in(`wrote').where(neq(`a'))

// how many articles did marko write with each 
//   coauthor?
g.V().has(`name',`marko').as(`a').out(`wrote').
 in(`wrote').where(neq(`a')).groupCount()
 
 // who are the 10 most central authors?
g.V().has(label,`person').
 repeat(out(`wrote').in(`wrote')).times(25).
 groupCount().order(local).by(valueDecr).limit(10)
\end{verbatim}
\end{small}

\subsection{Traverser Location and $\mu$-Superposition}

A classical random walk (traversal) is composed of a single walker (traverser) moving about the graph according to the instructions (steps) in $\Psi$ and the topology of $G$, where there will never be more than one traverser throughout the course of the computation (i.e. $|T| = 1$). However, in a quantum walk, there are two types of traversers: a single ``classical traverser" and set of ``quantum traversers." The classical traverser represents the initial state of the system. This state/traverser has a definite vertex location in a \textit{basis state} $|v\rangle$, where $v \in V$.\footnote{The notation $|x\rangle$ is known as ``bra(c)ket" notation and was introduced by Paul Dirac as it conveniently denotes row $\langle x|$ and column $|y\rangle$ vectors as well as operations such as inner products $\langle x|y\rangle$ and outer products $|y\rangle\langle x|$ \cite{bracket:dirac1939}. When represented programmatically, these vectors and matrices are arrays of complex numbers.} Once the initial state undergoes quantum processing, the classical traverser ($|T| = 1$) becomes a set of quantum traversers ($|T| > 1$). Moreover, each quantum traverser will yield more quantum traversers as more quantum operations are applied to the system, where, in the limit, $|T| \approx |V|$.\footnote{A potentially useful visualization is that of a quantum traverser at every vertex in the graph. These traversers can be thought of as a ``rubber sheet" across the vertices of the graph. As the quantum computation proceeds parts of the sheet will have more or less amplitude around 0, where there will never be more amplitude across the sheet than what was provided by the initial classical traverser (i.e.~the initial perturbation). Like other natural wave systems, quantum systems respect the conservation of energy.} After the initial step, the classical traverser is said to be in a \textit{superposition} of multiple definite vertex locations. Each location superposition is represented by a quantum traverser. The original classical traverser $t$ is defined by the quantum state vector
\begin{equation*}
|\mu(t)\rangle = c_1|v_1\rangle + c_2|v_2\rangle + \ldots + c_{|V|}|v_{|V|}\rangle, 
\end{equation*}
where $v_i \in V$ and $c_i \in \mathbb{C}$ is a complex number\footnote{A complex number is an element in $\mathbb{C}$. Every complex number is of the form
\begin{equation*}
c = a + bi,
\end{equation*}
where $c \in \mathbb{C}$, $a,b \in \mathbb{R}$, and $i = \sqrt{-1}$. The $a$ component is know as the real component and the $bi$ component is the imaginary component. Imaginary numbers ``rotate" a real number about the two-dimensional Argand plane because $i^0 = 1$, $i^1 = i$, $i^2 = -1$, $i^3 = -\sqrt{-1} = -i$, and $i^4 = 1$. Thus, multiplying a complex number by $i$ rotates it $90^\circ$. Multiplying a real number by $i^2$ rotates it $180^{\circ}$, turning a positive real to a negative real and vice versa. There exists a bijection that takes a complex number to a polar form $(\rho, \theta)$, where the magnitude $\rho = \sqrt{a^2 + b^2}$ and the phase $\theta = \text{tan}^{-1}\frac{b}{a}$.}  denoting the degree to which the original classical traverser is at the respective vertex in $V$. The classical traverser's location superposition is defined by a linear combination of the basis states along with a complex scalar and thus,
\begin{equation*}
|\mu(t)\rangle =
c_1 \begin{bmatrix}
1 \\
0 \\
\vdots \\
0 \\
\end{bmatrix} + 
c_2 \begin{bmatrix}
0 \\
1 \\
\vdots \\
0 \\
\end{bmatrix} + \ldots + 
c_{|V|} \begin{bmatrix}
0 \\
0 \\
\vdots \\
1 \\
\end{bmatrix} = 
\begin{bmatrix}
c_1 \\
c_2 \\
\vdots \\
c_{|V|} \\
\end{bmatrix},
\end{equation*}
where $[1,0,\ldots,0]^\top$ represents the vertex $v_1 \in V$.

It is important to emphasize that traverser location superposition is a native feature of the Gremlin traversal machine and language. A Gremlin traverser is a \textit{furcating traverser} in that whenever it meets a decision in $G$, the traverser will clone itself across all choices. For instance, if the traverser is at vertex $v \in V$ and $v$ has five outgoing \texttt{knows}-edges, then the step \texttt{out(`knows')} will yield five traversers located at the five adjacent \texttt{knows}-vertices of $v$. The reason for this is that Gremlin is a query language and the question being asked is ``return all the people that $v$ knows," not ``return one random person that $v$ knows." The latter result would be expected in a random walk, where $|T| = 1$ for the duration of the computation. 

In classical Gremlin, prior to the application of the constructs detailed in this article, when two or more traversers converge onto the same graph location, they will constructively interfere to generate a single traverser at that same graph location whose ``bulk" (i.e.~count) is the sum of all the bulks of the converging traversers. Bulking is not sufficient to enact a quantum walk. In a quantum walk, traversers must be able to constructively and destructively interfere with one another. As of the time of this writing, a traverser's bulk is represented by a 64-bit integer in $\mathbb{N}^+$ and thus, in order to represent ``bulks" with phases and amplitudes, the current bulk-construct must be abandoned (i.e.~disabled) and replaced by a complex vector in the traverser's sack called the traverser's spin.

\subsection{Traverser Spin and $\psi$-Superposition}

Quantum walks require both a location and a spin superposition to guarantee reversible, unitary evolution -- a primary requirement of quantum processing. Location superposition does not contain sufficient information to ensure that a complex wave dynamic is reversible. In Gremlin, a traverser has two ``locations." A graph location in $G$ ($\mu$) and traversal location in $\Psi$ ($\psi$). If only the graph location is in superposition, then information about the traversal location is lost. When a traverser has multiple options in $G$ (a topological branch), it must undergo graph location superposition. When a traverser has multiple options in $\Psi$ (a program branch), it must undergo \textit{spin superposition}.\footnote{Quantum walk theory uses the term ``spin superposition" to describe the various degrees of freedom in the walker's movement. However, in the terminology of graph traversals, a better term may be ``traversal superposition," where every branch in the program/traversal is a degree of freedom.} For each program branch option, there exists a complex number to represent it \cite{spin:nachtergaele2004}. For example, if the graph is a one-dimensional lattice (a line) and the traverser can either go left or right (\texttt{out(`left',`right')}), then the traverser $t$'s spin is encoded in its sack $\varsigma(t) \in \mathbb{C}^2$ which is composed of 2 complex numbers and is denoted
\begin{equation*}
|\varsigma(t)\rangle = c_1|\!\leftarrow\rangle + c_2|\!\rightarrow\rangle,
\end{equation*}
where
\begin{equation*}
c_1|\!\leftarrow\rangle + c_2|\!\rightarrow\rangle = 
c_1 \begin{bmatrix}
1 \\
0 \\
\end{bmatrix} + 
c_2 \begin{bmatrix}
0 \\
1 \\
\end{bmatrix} = 
\begin{bmatrix}
c_1 \\
c_2 \\
\end{bmatrix}.
\end{equation*}
The vectors $|\!\!\leftarrow\rangle$ and $|\!\!\rightarrow\rangle$ are the spin basis vectors and represent two orthogonal states that the classical traverser can be in. That is, when initialized or measured, the classical traverser will either be spinning left ($|\!\leftarrow\rangle$) or spinning right ($|\!\rightarrow\rangle$). However, while undergoing quantum processing, the quantum traversers can simultaneously have both a left spin and a right spin. The amount of spin in either direction is specified by $c_1 \in \mathbb{C}$ (left) and $c_2 \in \mathbb{C}$ (right). If there are four options in a traversal, then the traverser's spin vector would be $\varsigma(t) \in \mathbb{C}^4$.
 
A traverser's path is completely determined by its spin. If the traverser is on a line graph and its spin is $[1,0]^\top$, then the traverser will go left on the graph. If its spin is $[0,1]^\top$, then the traverser will go right on the graph.\footnote{The state $|\!\leftarrow\rangle$ represents the column vector $[1,0]^\top$. The state $\langle\leftarrow\!|$ represents the conjugate row vector $[1^*,0^*]$, where for a real number, the complex conjugate is the number itself and for a complex number $(a + bi)^* = a - bi$.} If its spin is in the superposition $[\frac{1}{\sqrt{2}},\frac{1}{\sqrt{2}}]^\top$, then the traverser will split itself and the left clone will go left with a spin of $[\frac{1}{\sqrt{2}},0]^\top$ and the right clone will go right with a spin of $[0,\frac{1}{\sqrt{2}}]^\top$. When two traversers exist at the same location in the graph and are within the same equivalence class $[t] = \{ t' \in T \;|\; \mu(t) = \mu(t') \}$\footnote{Traverser equivalence classes in Gremlin can be configured. For the concepts presented in this article, assume that two traversers map to the same equivalence class if their graph location is the same.}, the $[t]$-traversers will merge to a single traverser whose spin is determined using standard, pair-wise vector addition 
\begin{equation*}
\varsigma([t]) = \sum_{t' \in [t]} \varsigma(t').
\end{equation*}
It is the merging of traversers at a vertex location that yields the constructive and destructive wave interference patterns -- e.g. one traverser's spin may be positive ($i^0$) while another's may be negative ($i^2$). 

Given both location and spin superposition, the complete state of the classical traverser undergoing quantum processing is defined as the tensor product of the two superposition states
\begin{equation*}
|\mu(t)\rangle \otimes |\varsigma(t)\rangle \equiv |\mu(t) \otimes \varsigma(t)\rangle \equiv |\mu(t),\varsigma(t)\rangle,
\end{equation*}
where
\begin{equation*}
\begin{array}{llll}
|\mu(t),\varsigma(t)\rangle &=& 
 c_{1,1}|v_1,\leftarrow\rangle + c_{1,2}|v_1,\rightarrow\rangle \; & + \\ 
 && c_{2,1}|v_2,\leftarrow\rangle + c_{2,2}|v_2,\rightarrow\rangle \; & + \\ 
 && \ldots \; & + \\
 && c_{|V|,1}|v_{|V|},\leftarrow\rangle + c_{|V|,2}|v_{|V|},\rightarrow\rangle &.
\end{array}
\end{equation*}
A quantum system on a line graph can be described by a single complex vector in $\mathbb{C}^{2{|V|}}$. In quantum mechanics, this complex vector is known as the \textit{wavefunction} of the quantum system (i.e.~the classical traverser in superposition). However, in Gremlin, this representation is distributed across the graph, where each quantum traverser's location $\mu \in V$ represents the classical traverser's location superposition and each quantum traverser's spin is a length 2 complex vector $\varsigma \in \mathbb{C}^2$.\footnote{As of Gremlin 3.1.0, there is a tension between the traverser's bulk and its sack. A future version of Gremlin may generalize the bulk construct to any object that can be merged, split, and has a magnitude and thus, in such a situation, the traverser's spin would be encoded in its bulk.}

It is important to note that the wavefunction does not determine the location of the classical traverser as, in this representation, the traverser is a wave, not a particle. To transform the wave representation (which is encoded across all quantum traversers) into a particle representation, the system must be measured/observed. The act of measuring first transforms the wavefunction into a probability distribution, where for any vertex $v$, the probability of the classical traverser being at $v$ is defined by the square of the modulus of the total spin at that vertex.\footnote{The modulus of the vector $x$ is its magnitude $|x| = \sqrt{x_1^2 + x_2^2 + \ldots + x_n^2}$.} For instance, if the quantum traverser $t$ has $\mu(t) = v$ and $\varsigma(t) = [c_1,c_2]^\top$, then the probability of the traverser being at $v$ is the inner product
\begin{equation*}
 \langle\varsigma(t)|\varsigma(t)\rangle = 
\left[ c_1^*, c_2^* \right]
\begin{bmatrix}
c_1 \\
c_2 \\
\end{bmatrix} =  
 |c_1|^2 + |c_2|^2,
\end{equation*}
where for all quantum traversers on the line graph at any iteration $n$, 
\begin{equation*}
\sum_{t \in T} |\varsigma(t)_0|^2 + |\varsigma(t)_1|^2 = 1.
\end{equation*}
Thus, the initial classical traverser, when $n=1$ and $|T| = 1$, has an inner product equal to 1. Quantum processes respect the conservation of wave energy. This is analogous to other natural waves such as a water wave. The initial energy placed into the water will diffuse across the surface but the total energy of the system will never increase nor decrease (barring friction -- i.e. decoherence).

Once the wavefunction has been converted into a probability distribution, that distribution is sampled and the traverser is then localized to the respective sampled vertex. The wavefunction is said to \textit{collapse} to some classical, basis state vertex $v \in V$ and as originally, $|T| = 1$. Thus, a classical traverser undergoes quantum processing to yield numerous quantum traversers which are then sampled at some point in the future to yield the new location/spin of the original classical traverser.

A natural (non-simulated) quantum walk is both breadth-first and depth-first at the same time. It is breadth-first because at every step, all legal incident edges are traversed. It is depth-first because only one traverser (classical particle) is actually ever observed at the end of a path.\footnote{The de Broglie-Bohm pilot-wave interpretation of quantum mechanics states that there is only ever one particle (traverser) in a quantum process. A consequence of this line of reasoning is that there is no such thing as a particle superposition and a wavefunction collapse. The classical particle is simply guided by its own ``wave" (perturbation) in some yet unknown medium \cite{pilot:bohm1952}. Thus, the wave is doing a breadth-first walk, but the particle is doing a depth-first walk.} When simulating a quantum walk, the execution order must be breadth-first because each iteration needs to merge all co-located quantum traversers in order for wave interference to take place. In applied graph computing, this execution model is known as \textit{bulk synchronous parallel}, where a single traverser takes one step in the graph and does not proceed to do another until all other traversers have completed their current step \cite{bsp:valiant1990}. The benefit of this is that breadth-first execution can be easily parallelized/distributed. However, it requires more memory than depth-first as each traverser at every step must be represented. The trade-off between a breadth-first search and a depth-first search is typically a tradeoff between time (breadth-first) and space (depth-first). Interestingly, a natural quantum walk has the benefit of both without the drawback of either granted that the location of the observed particle is the correct answer to the search.

\subsection{Unitary Operators}

Quantum mechanics is about reversible computing. Reversibility is a constraint that significantly limits the types of operations that can be performed on a quantum system. It requires that every operation have an inverse that returns the system to its previous state. Moreover, that inverse operation is the operation's adjoint (i.e. complex conjugate).\footnote{The complex conjugate of a complex matrix is defined as
\begin{equation*}
\begin{bmatrix}
a + bi & c + di \\
e - fi & g - hi \\
\end{bmatrix}^\dagger = 
\begin{bmatrix}
a + bi & e + fi \\
c - di & g - hi \\
\end{bmatrix}.
\end{equation*}} Specifically, all operations must be unitary. A unitary matrix $U$ is any matrix that satisfies the relation $U^{\dagger}U = I$, where $U^{\dagger}$ is the complex conjugate of the matrix $U$. It is important to note that the set of all $n \times n$ unitary matrices forms an algebraic group, where matrix multiplication yields a unitary matrix ($A \cdot B$), each unitary matrix has an inverse ($A^\dagger$), and the multiplicative identity is the unitary identity matrix $I$. Furthermore, the tensor product of two unitary matrices is a unitary matrix ($A \otimes B$). 

When the quantum system's state is altered by $U$, its inner product remains 1. That is $(U | x\rangle)^\dagger U | x\rangle = \langle x |U^\dagger U|x\rangle = \langle x | I | x \rangle =  \langle x | x \rangle = 1$. All unitary operations are isometric in that they preserve this distance. However, if an operation is not unitary, it can still be ``quantum" in nature, but it will distort the geometry of the system. This is known as \textit{decoherence} and decoherence is fundamentally a process of information loss and thus, irreversibility. 

A quantum walk has two states in superposition -- location and spin. There are two unitary operations that act on these states: a ``coin"-operator and a ``shift"-operator. The coin-operator transforms a traverser's spin vector $\varsigma(t)$ into a new vector which is then propagated piecewise by the shift-operator to update $\mu(t)$. In a one-dimensional line graph, every coin-operator is a $2\times2$ unitary matrix in $\mathbb{C}^{2\times2}$. Two typical $2\times2$ coin operators used in quantum walks are the unbalanced Hadamard coin
\begin{equation*}
H = \frac{1}{\sqrt{2}}\begin{bmatrix}
1 & 1 \\
1 & -1 \\
\end{bmatrix}
\end{equation*}
and the balanced coin
\begin{equation*}
Y = \frac{1}{\sqrt{2}}
\begin{bmatrix}
1 & i \\
i & 1 \\
\end{bmatrix},
\end{equation*}
where $H^{\dagger}H = I$ and $Y^{\dagger}Y = I$. The altered spin vector of traverser $t$ is thus $H | \varsigma(t) \rangle$. Once the spin has been updated by the coin-operator, the traverser's $\mu(t)$ location is updated by the shift-operator
\begin{equation*}
S|\mu(t),\varsigma(t)\rangle = |\mu(t) - 1,[\varsigma(t)_0,0]^\top\rangle + |\mu(t) +1,[0,\varsigma(t)_1]^\top\rangle,
\end{equation*}
where $\mu(t) - 1$ ($\mu(t) + 1)$ is the vertex to the left (right) of $\mu(t)$. In other words, the left-traverser moves left one vertex and has its right spin component set to 0. Similarly, the right-traverser moves to the right one step and has its left spin component set to 0. The complete unitary operation of the entire system (not just a single quantum traverser) is defined as
\begin{equation*}
U = S \cdot (I \otimes C),
\end{equation*}
where $I \in \{0,1\}^{|V| \times |V|}$ is the identity matrix. The formulation implies that the coin operator $C$ is ``copied" to each vertex in the graph and the shift operator then propagates spin accordingly. However, in Gremlin, this single matrix representation is distributed across the graph where the logic of the coin- and shift-operators are evaluated by each quantum traverser. At each step, each quantum traverser evaluates $C$ and then $S$ to yield a new set of quantum traversers. Finally, because unitary matrices form an algebraic group with matrix multiplication, $U^n$ is unitary and it will iterate this process $n$-times, propagating the wavefunction $n$-steps on the graph. Thus, an $n$-step quantum walk is unitary.

\section{Quantum Walk Experiments}

A quantum computation starts with a classical traverser in a basis state representing the initial state of the quantum process. The classical traverser is then operated on by a unitary operator. This operation puts the system into a superposition. When the system is measured, a classical traverser is yielded whose location is determined by the wavefunction of the quantum system. In the natural-world, the wavefunction can never be directly observed, only a resultant basis state is observed after measurement. However, when simulating a quantum system, the wavefunction forms the primary data structure of the computation and thus, it is subject to runtime analysis. This section will present a collection of quantum experiments on simple graph structures to demonstrate the representation of quantum walks in the Gremlin language.

\subsection{Classical Walk on a Line Graph}

The simplest walk to execute is one that takes place on a one-dimension lattice graph $G$ (a line graph) with no boundaries. To demonstrate such a walk, a line graph with $|V|=100$ is constructed, where each vertex has one outgoing \texttt{left}-edge and one outgoing \texttt{right}-edge. In order to ensure that no boundaries are touched, the walk will start at $v_{50} \in V$ and iterate for 50 iterations. To demonstrate the difference between a classical walk and a quantum walk, this section will first present the traversal and results of a classical walk using Gremlin. A classical random walk, where $|T| = 1$ at every iteration, can be evaluated using the \texttt{sample}-step.
\begin{small}
\begin{verbatim}
g.V(50).
 repeat(out(`left',`right').sample(1)).times(50)
\end{verbatim}
\end{small}
A traverser is placed on vertex 50 via \texttt{V(50)} and then for 50-iterations, the traverser will go both left and right on the line graph. However, while two traversers are created at each iteration (one left and one right), one of the two will be filtered by \texttt{sample(1)}. This ensures that only a single traverser exists at each time step and that that traverser is a ``random walker." Given that Gremlin natively supports traverser location superposition, the long run behavior of a classical random walk can be derived from the normalization of a non-\texttt{sample(1)} traversal's traverser counts (bulks) across $G$ (i.e.~frequency distribution).
\begin{small}
\begin{verbatim}
g.V(50).repeat(out(`left',`right')).times(50)
\end{verbatim}
\end{small}
The vertex frequency distribution can be generated by postfixing a \texttt{groupCount()}-step. This step returns a \texttt{Map<Vertex,Long>} denoting how many traversers are located at each vertex location.
\begin{small}
\begin{verbatim}
g.V(50).
 repeat(out(`left',`right')).times(50).
  groupCount()
\end{verbatim}
\end{small}

The frequency distribution of the first 5 steps of the classical walk is provided in Table \ref{tab:classic-frequency-line}. Note that at step 3, the two 1s in step 2 split left and right and merge at vertex 50 to constructively interfere to create a count of 2. Again, in a classical walk, only constructive/additive interference ever occurs.\footnote{It is interesting to note that a traverser's bulk is represented by a 64-bit integer. Programmatically, if the bulk is larger than what a 64-bit integer can represent, then a number overflow occurs and the bulk becomes negative and thus, destructive interference occurs upon merging a non-overflow with an overflow bulk. A speculation is that the natural world is in fact premised on bulking/counts, but due to numeric precision issues (fidelity), the wave top crests and becomes negative. In fact, similar results using overflow on bounded precision bulks (16-bit, 32-bit, etc.) can effect results similar to the quantum experiments to follow.} These counts are normalized and presented in Table \ref{tab:classic-line}. Note that Table \ref{tab:classic-line} also shows the probability distribution at iteration $50$. At iteration $50$, the number of traversers at vertex 50 is $126,410,606,437,752$ ($\sim126$ trillion traversers) and the total number of represented traversers across the whole graph is $1,125,899,906,842,624$ ($\sim1$ quadrillion traversers).\footnote{Gremlin uses frequentist ``bulking" when merging traversers in the same equivalence class $[t]$. Thus, while 1 quadrillion traversers are represented, they are not individually enumerated.}
\begin{center}
\begin{table}
\begin{tabular}{ |l|c|c|c|c|c|c|c|c|c|c|c| } \hline
$n/V$ & \ldots & 46 & 47 & 48 & 49 & 50 & 51 & 52 & 53 & 54 & \ldots \\ \hline\hline
1 & \ldots & 0 & 0 & 0 & 0 & 1 & 0 & 0 & 0 & 0 & \ldots\\ \hline
2 & \ldots & 0 & 0 & 0 & 1 & 0 & 1 & 0 & 0 & 0 & \ldots\\ \hline 
3 & \ldots & 0 & 0 & 1 & 0 & 2 & 0 & 1 & 0 & 0 & \ldots\\ \hline 
4 & \ldots & 0 & 1 & 0 & 3 & 0 & 3 & 0 & 1 & 0 & \ldots\\ \hline 
5 & \ldots & 1 & 0 & 4 & 0 & 6 & 0 & 4 & 0 & 1 & \ldots\\ \hline 
\end{tabular}
\caption {The frequency distribution of traversers at each iteration in a classical walk on a line graph.}\label{tab:classic-frequency-line} 
\end{table}
\end{center}
\begin{center}
\begin{table}
\begin{tabular}{ |l|c|c|c|c|c|c|c|c|c|c|c| } \hline
$n/V$ & \ldots & 46 & 47 & 48 & 49 & 50 & 51 & 52 & 53 & 54 & \ldots \\ \hline\hline
1 & \ldots & 0 & 0 & 0 & 0 & 1 & 0 & 0 & 0 & 0 & \ldots\\ \hline
2 & \ldots & 0 & 0 & 0 & $\frac{1}{2}$ & 0 & $\frac{1}{2}$ & 0 & 0 & 0 & \ldots\\ \hline 
3 & \ldots & 0 & 0 & $\frac{1}{4}$ & 0 & $\frac{1}{2}$ & 0 & $\frac{1}{4}$ & 0 & 0 & \ldots\\ \hline 
4 & \ldots & 0 & $\frac{1}{8}$ & 0 & $\frac{3}{8}$ & 0 & $\frac{3}{8}$ & 0 & $\frac{1}{8}$ & 0 & \ldots\\ \hline 
5 & \ldots & $\frac{1}{16}$ & 0 & $\frac{4}{16}$ & 0 & $\frac{6}{16}$ & 0 & $\frac{4}{16}$ & 0 & $\frac{1}{16}$ & \ldots\\ \hline 
\ldots & \ldots & \ldots & \ldots & \ldots & \ldots & \ldots & \ldots & \ldots & \ldots & \ldots & \ldots\\ \hline 
50 & \ldots & 0.096 & 0 & 0.108 & 0 & 0.112 & 0 & 0.108 & 0 & 0.096 & \ldots\\ \hline 
\end{tabular}
\caption {The probability distribution of a traverser's vertex location at each iteration in a classical random walk.}\label{tab:classic-line} 
\end{table}
\end{center}

\subsection{Unbalanced Quantum Walk on a Line Graph}

A classical walk can be compactly described in Gremlin. On the other hand, a quantum walk in Gremlin is more complex given that it requires that each step modulate the traverser's spin and projects that spin component-wise to the adjacent left and right vertices accordingly. This complexity exists because spin superposition, unlike location superposition, is not natively supported in Gremlin. Before presenting the Gremlin quantum walk traversal, three functions are defined. First, the \texttt{merge} function is responsible for constructive and destructive wave interference. It merges the sacks (spins) of any two traversers at the same vertex location via pairwise vector addition.
\begin{small}
\begin{verbatim}
merge = { a,b -> [a[0] + b[0], a[1] + b[1]] }.
\end{verbatim}
\end{small}
Second, the Hadamard coin,
\begin{equation*}
H = \frac{1}{\sqrt{2}}\begin{bmatrix}
1 & 1 \\
1 & -1 \\
\end{bmatrix},
\end{equation*}
is computed with the \texttt{hadamard} function, where the $b$-argument is \texttt{null}.\footnote{All \texttt{sack}-steps take a binary function. However, if the second argument is not required because the function itself contains the full logic for mutating the first argument, then it can be safely ignored.}
\begin{small}
\begin{verbatim}
hadamard = { a,b -> 
              [(1/Math.sqrt(2)) * (a[0] + a[1]),   
               (1/Math.sqrt(2)) * (a[0] - a[1])]}.
\end{verbatim}
\end{small}
Lastly, the \texttt{shift} function forces a traverser's spin vector into a basis state by either setting the left- or right-spin component to 0, where $b=[1,0]$ will project the left-spin component and  $b=[0,1]$ will project the right-spin component.  
\begin{small}
\begin{verbatim}
shift = { a,b -> [a[0] * b[0], a[1] * b[1]] }.
\end{verbatim}
\end{small}

A quantum walk is started with a classical traverser located at vertex 50 with a spin in the basis state $[1,0]^\top$. In Dirac notation, this is the state $|v_{50},\leftarrow \rangle$.\footnote{In Gremlin 3.1.0, a \texttt{withBulk(false)} parameterization is required after \texttt{g} to tell the Gremlin machine to use a traverser's sack (not its bulk) as the determinant of the traverser's magnitude/count. This parameterization is left out of the presented Gremlin quantum examples. As stated previously, a future version of Gremlin may generalize the bulk construct to be anything that can be ``split" and ``merged" and thus, at that point, the traverser's spin will be represented in the bulk of the traverser, not its sack.}
\begin{small}
\begin{verbatim}
g.withSack([1,0],merge).V(50).
 repeat(
  sack(hadamard). 
  union(
   sack(shift).by(constant([1,0])).out(`left'),
   sack(shift).by(constant([0,1])).out(`right'))).
 times(50)
\end{verbatim}
\end{small}
The \texttt{withSack([1,0],merge)} parameterization defines the initial state of the traverser's sack along with the merge function to use when two traversers meet at the same vertex location. The traverser is then placed at vertex 50 via \texttt{V(50)}. Finally, 50 iterations are executed, where for each iteration, all traverser sacks are evolved by the unitary \texttt{hadamard} function and then two traversers are created each with the original traverser's left (right) sack component zeroed out by the \texttt{shift} function. These traverser children are then propagated left or right on the line graph accordingly.

\begin{figure}[h!]
	\centering
	\includegraphics[width=0.49\textwidth]{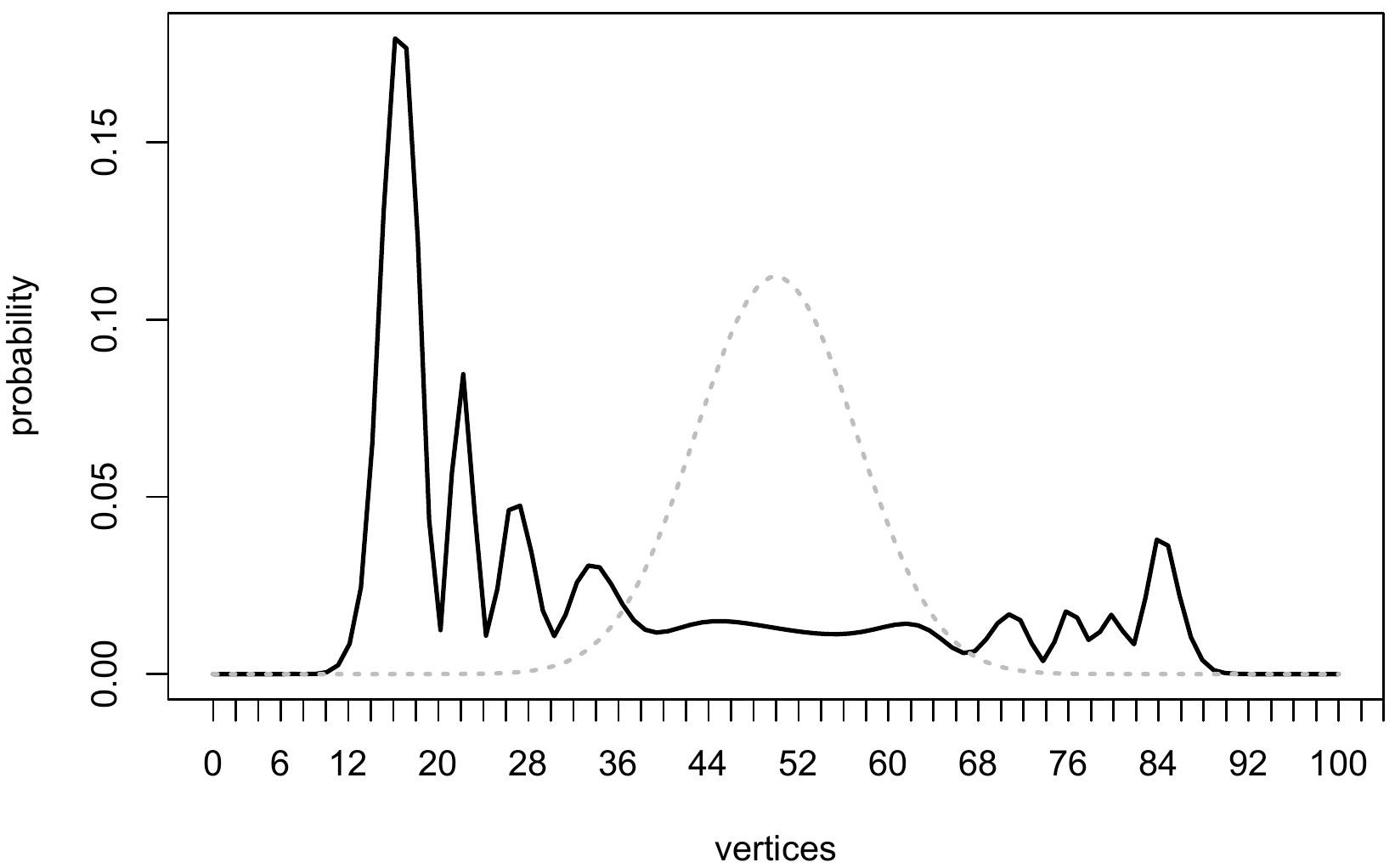}
	 \caption{\label{fig:hadamard-line}The probability of locating a classical walker (gray dashed line) and a Hadamard-coin quantum walker (black solid line) at a particular vertex on a 100 vertex line graph after 50 iterations.}
\end{figure}

In order to yield the probability that the classical traverser will be at some vertex in $V$, the wavefunction (quantum traversers) must be turned into a probability distribution. This is accomplished by grouping and norming the sack of each traverser. If
\begin{small}
\begin{verbatim}
norm = { sack -> 
 Math.pow(sack.get()[0],2) + 
 Math.pow(sack.get()[1],2) 
}
\end{verbatim}
\end{small}
then the probability distribution is computed using
\begin{small}
\begin{verbatim}
g.withSack([1,0],merge).V(50).
 repeat(
  sack(hadamard). 
  union(
   sack(shift).by(constant([1,0])).out(`left'),
   sack(shift).by(constant([0,1])).out(`right'))).
 times(50).group().by().by(sack().map(norm)).
\end{verbatim}
\end{small}
The \texttt{group}-step takes two \texttt{by}-modulators. The first is the group ``key" where \texttt{by()} is the identity and thus, the vertex location. The second is the modulus squared of the spin at that vertex. The probability distribution of the traverser's location for the first five and last iteration is presented in Table \ref{tab:hadamard-line}. Note that at step 4, unlike a classical walk, destructive interference at other vertices increases the probability of locating the traverser at vertex 49. The traverser moving right is $180^{\circ}$ out of phase with the traverser moving left.\footnote{The complex number $1 + 0i$ (1) is $180^{\circ}$ out of phase with $0 + i^2$ (-1). The polar form $(\rho,\theta)$ of these two complex numbers is $(1,0)$ and $(1,180)$, respectively.}
\begin{center}
\begin{table}
\begin{tabular}{ |l|c|c|c|c|c|c|c|c|c|c|c| } \hline
$n/V$ & \ldots & 46 & 47 & 48 & 49 & 50 & 51 & 52 & 53 & 54 & \ldots \\ \hline\hline
1 & \ldots & 0 & 0 & 0 & 0 & 1 & 0 & 0 & 0 & 0 & \ldots\\ \hline
2 & \ldots & 0 & 0 & 0 & $\frac{1}{2}$ & 0 & $\frac{1}{2}$ & 0 & 0 & 0 & \ldots\\ \hline 
3 & \ldots & 0 & 0 & $\frac{1}{4}$ & 0 & $\frac{1}{2}$ & 0 & $\frac{1}{4}$ & 0 & 0 & \ldots\\ \hline 
4 & \ldots & 0 & $\frac{1}{8}$ & 0 & $\frac{5}{8}$ & 0 & $\frac{1}{8}$ & 0 & $\frac{1}{8}$ & 0 & \ldots\\ \hline 
5 & \ldots & $\frac{1}{16}$ & 0 & $\frac{10}{16}$ & 0 & $\frac{2}{16}$ & 0 & $\frac{2}{16}$ & 0 & $\frac{1}{16}$ & \ldots\\ \hline 
\ldots & \ldots & \ldots & \ldots & \ldots & \ldots & \ldots & \ldots & \ldots & \ldots & \ldots & \ldots\\ \hline 
50 & \ldots & 0.015 & 0 & 0.014 & 0 & 0.013 & 0 & 0.012 & 0 & 0.011 & \ldots\\ \hline 
\end{tabular}
\caption {The probability distribution of the traverser's vertex location at each iteration using an initial spin of $[1,0]^\top$ and the $H$ Hadamard coin.} \label{tab:hadamard-line} 
\end{table}
\end{center}
The probability distribution at iteration 50 for both classical and quantum Hadamard walks is diagrammed in Figure \ref{fig:hadamard-line}. Given that the Hadamard is a biased coin, the traverser's location is biased to the left. Moreover, the traverser is less likely to be located at the center of the line which is contrary to what is expected from the classical walk. For a classical random walk, after $n$-steps, the traverser will most likely be at a vertex $\sqrt{n}$-step away from $v_{50}$. In a quantum walk, the traverser will most likely be at a vertex $n$-steps away from $v_{50}$. This feature makes quantum walks interesting as an algorithmic technique as, with the correct coin and initial state, quantum walks can be self-avoiding in search of novel, less reverberant/noisy areas of the graph \cite{quantapps:ambainis2003}.

The previous traversal generated a probability distribution from the wavefunction amplitudes. However, the final step is to ultimately collapse the wavefunction to a basis state. This is accomplished by sampling the probability distribution. Thus, the full quantum traversal, from classical start state to classical end state is below.
\begin{small}
\begin{verbatim}
g.withSack([1,0],merge).V(50).
 repeat(
  sack(hadamard). 
  union(
   sack(shift).by(constant([1,0])).out(`left'),
   sack(shift).by(constant([0,1])).out(`right'))).
 times(50).
  group().by().by(sack().map(norm)).
   unfold().sample(1).by{kv -> kv.value}.select(keys)
\end{verbatim}
\end{small}
The last line of the traversal samples the \texttt{group}-step probability distribution. It \texttt{unfold}s the key/values pairs of the \texttt{Map<Vertex,Double>} and samples 1 key/value pair where the probability distribution is over the set of all key/value pair values. Given that the basis state is a vertex and not a probability, the key of the sampled key/value pair is projected with \texttt{select(keys)}. In review, the first line of the traversal creates the classical initial state (naturally observable), the \texttt{repeat(...).times(50)} lines are the evolution of the quantum state (naturally unobservable), and the last two lines represent the wavefunction collapse (unobservable) which ultimately yields a single classical traverser at a vertex (observable).

Finally, it is important to note that the traversal, prior to wavefunction collapse, is unitary. This can be demonstrated by iterating for 50 steps and then iterating the inverse 50 steps. After doing so, vertex $50$ will have a single traverser with spin $[1,0]^\top$ and probability $|1|^2 + |0|^2 = 1.0$.\footnote{In practice, a complex number is represented by two 64-bit floating point numbers. Numerous operations on such numbers typically incur floating point errors and thus, a perfect $1.0$ reconstruction of the classical traverser's spin is unlikely. Usually, results are of the form $0.999999997$.} That is, the classical traverser can be reconstructed by running the quantum process in reverse.
\begin{small}
\begin{verbatim}
g.withSack([1,0],merge).V(50).
 repeat(
  sack(hadamard). 
  union(  
   sack(shift).by(constant([1,0])).out(`left'),
   sack(shift).by(constant([0,1])).out(`right'))).
 times(50).
 repeat(
  union(
   sack(shift).by(constant([1,0])).in(`left'),
   sack(shift).by(constant([0,1])).in(`right')).
  sack(hadamard)).
 times(50)
\end{verbatim}
\end{small}

\subsection{Balanced Quantum Walk on a Line Graph} 

\begin{figure}[h!]
	\centering
	\includegraphics[width=0.49\textwidth]{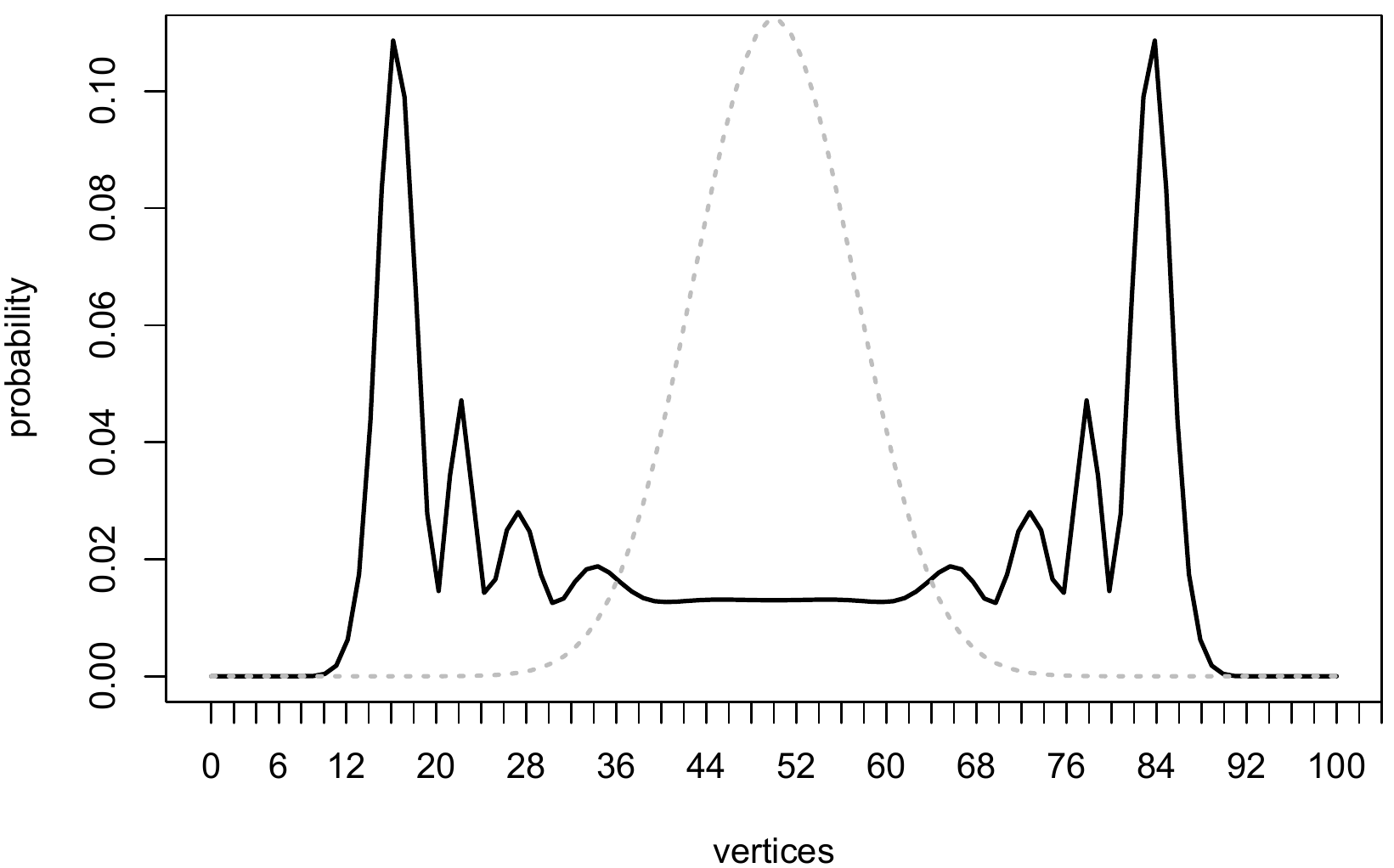}
	 \caption{\label{fig:balanced-line}The probability of locating a classical walker (gray dashed line) and an unbiased $Y$-coin quantum walker (black solid line) at a particular vertex on a 100 vertex line graph after 50 steps.}
\end{figure}

The Hadamard coin $H$ is useful as an introductory example because it is a coin in $\mathbb{R}^{2\times2}$. However, $\mathbb{R} \subset \mathbb{C}$ and thus, for arbitrary complex coins and spins, the previous quantum walk traversal must be generalized to support complex numbers. The Apache Commons Math3 library's \texttt{org.apache.commons.math3.complex.Complex} can be imported into Gremlin using the \texttt{:import} command.\footnote{The Apache Commons Math3 \texttt{Complex} number is encoded as a two dimensional \texttt{double} array where the first component of the array is the real component ($a$) and the second is the real number in the imaginary component ($b$), where $a + bi$. Thus $i^0 = (1,0)$, $i = (0,1)$, $i^2 = (-1,0)$, $i^3 = (0,-1)$ and $i^4=(1,0)$, etc.} From there, if
\begin{small}
\begin{verbatim}
zero = Complex.valueOf(0,0)
one = Complex.valueOf(1,0)
i = Complex.valueOf(0,1)
\end{verbatim}
\end{small}
then the $2 \times 2$ balanced unitary coin
\begin{equation*}
Y = \frac{1}{\sqrt{2}}
\begin{bmatrix}
1 & i \\
i & 1 \\
\end{bmatrix}
\end{equation*}
is denoted
\begin{small}
\begin{verbatim}
balanced = { a,b -> 
  [(1/Math.sqrt(2)) * a[0].add(i.mult(a[1])),   
   (1/Math.sqrt(2)) * i.mult(a[0]).add(a[1])] }.
\end{verbatim}
\end{small}
Furthermore, the \texttt{merge} function must be redefined to support pairwise complex vector addition.\footnote{The binary operators \texttt{+} and \texttt{*} can only be used with standard Java primitives like \texttt{long}, \texttt{double}, etc. To add and multiply a \texttt{Complex} number, the respective \texttt{Complex} methods must be used.}
\begin{small}
\begin{verbatim}
merge = { a,b -> [a[0].add(b[0]),a[1].add(b[1])] }
\end{verbatim}
\end{small}
The balanced coin $Y$ uses the imaginary number $i$ to rotate the traverser's spin by $90^{\circ}$ on each iteration. The following Gremlin traversal leverages complex numbers and the balanced $Y$-coin.
\begin{small}
\begin{verbatim}
g.withSack([one,zero],merge).V(50).
 repeat(
  sack(balanced). 
  union(
   sack(shift).by(constant([one,zero])).out(`left'),
   sack(shift).by(constant([zero,one])).out(`right')
 )).times(50)
\end{verbatim}
\end{small}
The probability distribution generated by the $Y$-traversal is diagramed in Figure \ref{fig:balanced-line} and the evolution of the probability distribution is provided in Table \ref{tab:balanced-line}. The $Y$-coin is balanced and thus, both the left and right components of the distribution centered at vertex $50$ are identical. However, like $H$, the probability of locating the traverser at vertex 50 is minimal compared to the vertices on the line's ends.

\begin{center}
\begin{table}
\begin{tabular}{ |l|c|c|c|c|c|c|c|c|c|c|c| } \hline
$n/V$ & \ldots & 46 & 47 & 48 & 49 & 50 & 51 & 52 & 53 & 54 & \ldots \\ \hline\hline
1 & \ldots & 0 & 0 & 0 & 0 & 1 & 0 & 0 & 0 & 0 & \ldots\\ \hline
2 & \ldots & 0 & 0 & 0 & $\frac{1}{2}$ & 0 & $\frac{1}{2}$ & 0 & 0 & 0 & \ldots\\ \hline 
3 & \ldots & 0 & 0 & $\frac{1}{4}$ & 0 & $\frac{1}{2}$ & 0 & $\frac{1}{4}$ & 0 & 0 & \ldots\\ \hline 
4 & \ldots & 0 & $\frac{1}{8}$ & 0 & $\frac{3}{8}$ & 0 & $\frac{3}{8}$ & 0 & $\frac{1}{8}$ & 0 & \ldots\\ \hline 
5 & \ldots & $\frac{1}{16}$ & 0 & $\frac{6}{16}$ & 0 & $\frac{2}{16}$ & 0 & $\frac{6}{16}$ & 0 & $\frac{1}{16}$ & \ldots\\ \hline 
\ldots & \ldots & \ldots & \ldots & \ldots & \ldots & \ldots & \ldots & \ldots & \ldots & \ldots & \ldots\\ \hline 
50 & \ldots & 0.013 & 0 & 0.013 & 0 & 0.013 & 0 & 0.013 & 0 & 0.013 & \ldots\\ \hline 
\end{tabular}
\caption {The probability distribution of the traverser's vertex location at each iteration using an initial spin of $[1,0]^\top$ and the balanced $Y$-coin.} \label{tab:balanced-line} 
\end{table}
\end{center}

\subsection{Quantum Walk on a Line Graph with Boundaries}

\begin{figure}[h!]
	\centering
	\includegraphics[width=0.49\textwidth]{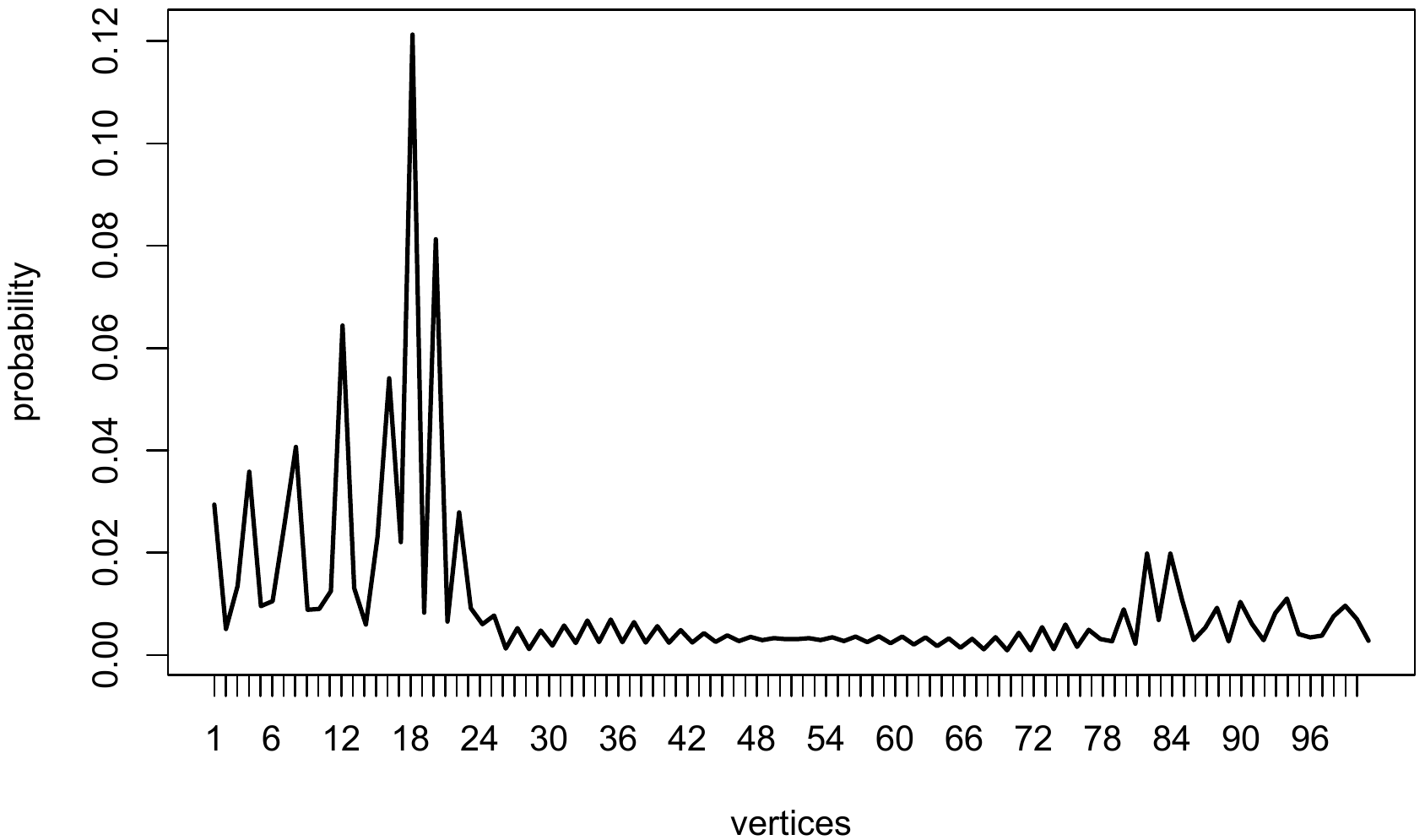}
	 \caption{\label{fig:hadamard-boundary-line}The probability of locating a Hadamard-coin quantum walker with boundary support at a particular vertex on a 100 vertex line graph after 100 iterations.}
\end{figure}

The previous simulations were run with an initial traverser placed at the middle of the line and using $|V| / 2$ number of iterations (e.g.~50 steps for a 100 vertex line graph). This ensured that the traversers did not try and move beyond the boundary vertices $v_1$ and $v_{|V|}$. However, it is rarely the case in real-world situations that the quantum process will be conveniently confined to an unbounded space. In the case where a traverser with spin $[c_1,c_2]^\top$ reaches $v_1$, the unitary operator for the next step generates two traversers where the right-traverser is placed at $v_2$ with spin $[0,c_2]^\top$ (typical) and the left-traverser remains at vertex $v_1$ with spin $[0,c_1]^\top$ (atypical). In words, the left traverser remains at $v_1$ with its left-spin component now being its right-spin component (i.e.~reflection). If
\begin{small}
\begin{verbatim}
reflect = { a,b -> [a[1], a[0]] }
\end{verbatim}
\end{small}
then a unitary reverberant quantum walk using $H$ for 100 iterations is defined in Gremlin as
\begin{small}
\begin{verbatim}
g.withSack([1,0],sackSum).V(50).
 repeat(
  sack(hadamard). 
  union(
   choose(out(`left').count().is(gt(0)),
    sack(shift).by(constant([1,0])).out(`left'),
    sack(shift).by(constant([1,0])).sack(reflect)),
   choose(out(`right').count().is(gt(0)), 
    sack(shift).by(constant([0,1])).out(`right'),
    sack(shift).by(constant([0,1])).sack(reflect)))).
 times(100).
\end{verbatim}
\end{small}
When there are only two options in a \texttt{choose}-step, the meaning of \texttt{choose} is ``if-then-else." If there is a left (right) vertex, then the traverser is cloned and shifts as previous. Else, if no such vertex exists (i.e.~a boundary is reached), then the traverser's spin is shifted and then reflected. There are four branching options in the boundary-aware traversal above -- two unions each with two chooses ($2 * 2 = 4$). However, because the \texttt{choose}-step branches are selective and based on the state of the quantum traverser, only one branch will ever be taken for each quantum traverser. Thus, the sack still only needs to be in $\mathbb{C}^2$. The probability distribution generated from the traversal above is plotted in Figure \ref{fig:hadamard-boundary-line}.

The reflection operation remains unitary. The original \S 2 specification of the unitary operator of a quantum walk was $U = S \cdot (C \otimes I)$, where the coin $C$ is ``copied" to each vertex via the tensor product of the identity matrix $I$. It is not necessary that the same coin $C$ be used at each vertex in $V$. In fact, as long as the coin used is unitary, there can be a heterogenous set of coins used in a quantum walk \cite{hybrid:karafyllidis2014}. This is analogous to how quantum circuits work. The graph is a collection of unitary \textit{quantum gates} (vertices) connected to one another via quantum wires (edges). Each gate can perform a different unitary operation on the \textit{qubit} (walker). There are two points to be made. First, the \texttt{reflect} operation is simply a different unitary operation being used at the boundary vertices. Second, quantum graph walks are sufficiently expressive for universal quantum computing \cite{universal:childs2009} and thus, Gremlin can serve as a general purpose quantum programming language. 

\subsection{A Quantum Walk on a Two-Dimensional Lattice with Two Slits}

In 1802, Thomas Young published the results of a light experiment known today as the ``double-slit experiment \cite{light:young1802}." In this experiment, a light source emits light at a screen. The screen has two slits in it. The light that makes it through the slits ultimately ends up being registered by a light sensitive film. The results of this experiment demonstrated that light behaves as a wave because the film showed interference patterns typical of wave mechanics. However, in 1932, Sir Geoffrey Ingram Taylor repeated the experiment, where instead of the light source emitting a constant stream of light, it emitted a single quanta of light known as a photon (a ``feeble" amount of light) \cite{light:taylor1909}. Unexpectedly, the same interference pattern emerged in the long run -- a single photon can interfere with itself! From the work of both Young and Taylor, light is now understood as being both a particle (at emission from the light source and absorption by the film) and a wave (while in quantum superposition between the start and end states). In other words, light is a particle when in a basis state and is a wave when in a superposition of the basis states.

\begin{figure}[h!]
	\centering
	\includegraphics[width=0.35\textwidth]{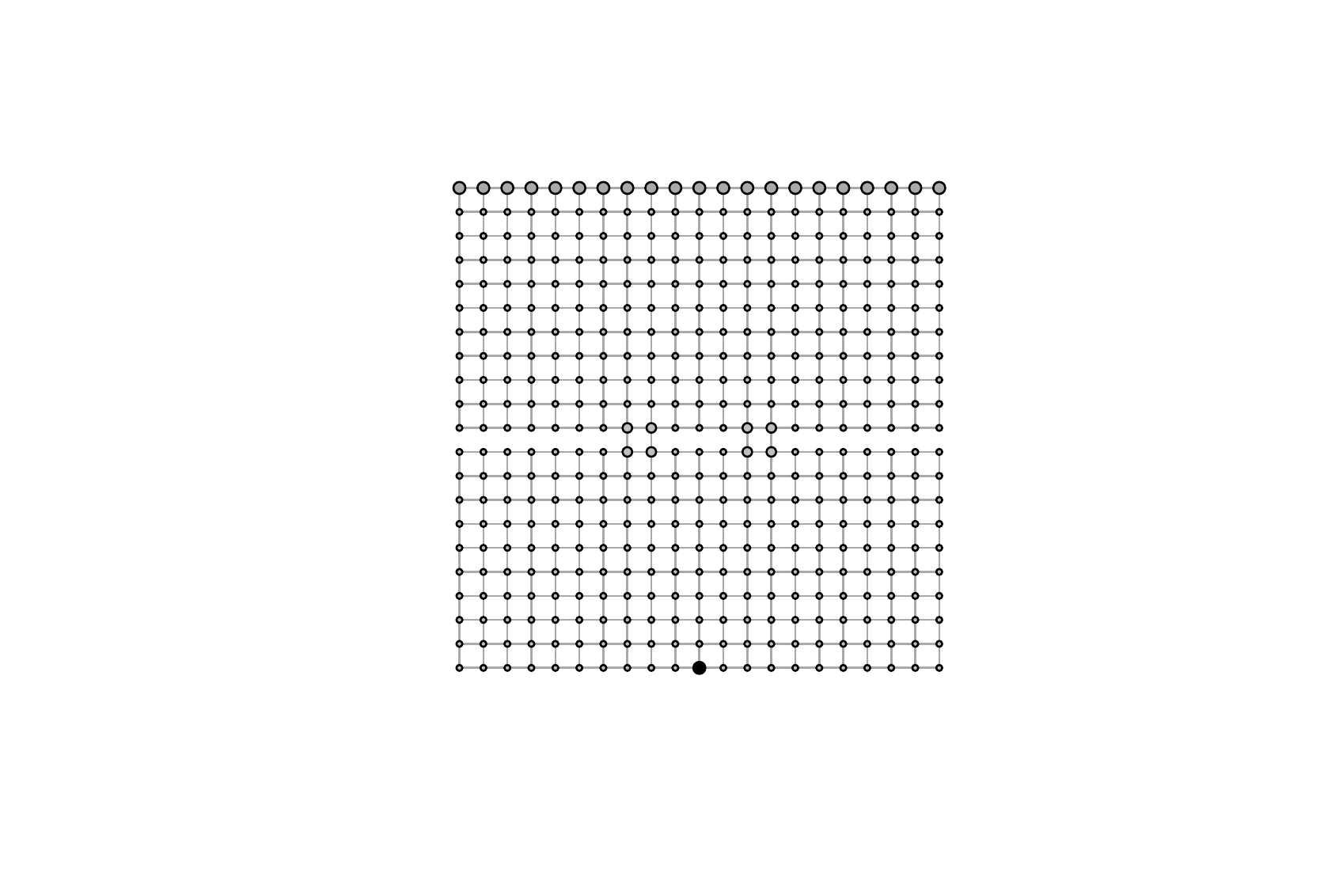}
	 \caption{\label{fig:double-slit-lattice}The two-dimensional lattice used to simulate the double-slit screen experiment. The black bottom center vertex is provided a single classical traverser in the basis state $[0,0,1,0]^\top$ (i.e. spin up). Only four vertices in the $10^{\text{th}}$ and $11^\text{th}$ rows of the lattice have edges and they represent the two 2-vertex width slits in the screen. Finally, the dark gray vertices at the top of the lattice represent the light sensitive film which will ultimately measure the probability distribution.}
\end{figure}

The double-slit experiment can be repeated using Gremlin. It requires a ``space" (a two-dimensional lattice), a slit screen (vertices with and without edges), and a light sensitive film (the back row vertices on the lattice) \cite{doubleslit:oliverira2007,twoquantum:oliverira2006}. Figure \ref{fig:double-slit-lattice} visualizes the two-dimensional lattice, where the bottom black vertex is the initial location of the classical traverser (the light source), the four vertices with edges at the $10^\text{th}$ and $11^\text{th}$ rows are the two slits, and the 20 vertices at the top of the lattice represent the film.

Given that a two dimensional lattice has 4 directions (left, right, up, and down) and barrier conditions, the coin must be a $4 \times 4$ unitary operator that supports reflection. The balanced Grover coin is a $4 \times 4$ unitary operator that will rotate the traverser $180^\circ$ ($i^2 = -1$) in its current trajectory/spin and only $\frac{1}{2}$-scale it in the other directions. The Grover coin is defined as
\begin{equation*}
R = \frac{1}{2}
\begin{bmatrix}
-1 & 1 & 1 & 1 \\
1 & -1 & 1 & 1 \\
1 & 1 & -1 & 1 \\
1 & 1 & 1 & -1 \\
\end{bmatrix},
\end{equation*}
where, in Gremlin, the $R$ operation is computed using
\begin{small}
\begin{verbatim}
grover = { a,b -> 
  [0.5 * (-a[0] + a[1] + a[2] + a[3]), 
   0.5 * ( a[0] - a[1] + a[2] + a[3]), 
   0.5 * ( a[0] + a[1] - a[2] + a[3]),
   0.5 * ( a[0] + a[1] + a[2] - a[3])]
}.
\end{verbatim}
\end{small}
Given that there are four traversal branches, \texttt{merge}, \texttt{shift}, and \texttt{reflect} must be defined accordingly, where for \texttt{reflect}, ``ud" represents reflection on the up-down axis and ``lr" represents reflection on the left-right axis.
\begin{small}
\begin{verbatim}
merge = { a,b -> 
 [a[0] + b[0], a[1] + b[1], 
  a[2] + b[2], a[3] + b[3]] 
} 

shift = { a,b ->  
 [a[0] * b[0], a[1] * b[1], 
  a[2] * b[2], a[3] * b[3]]
}

reflect = {a,b -> b == `ud' ? 
 [a[0],a[1],a[3],a[2]] : 
 [a[1],a[0],a[2],a[3]] 
}
\end{verbatim}
\end{small}
A two-dimensional lattice walk with the Grover coin $R$ is expressed below where vertex 10 is the bottom center vertex (dark black vertex in Figure \ref{fig:double-slit-lattice}).
\begin{small}
\begin{verbatim}
g.withSack([0,0,1,0],merge).V(10).
 repeat(
  sack(grover).
  union(
   choose(out(`left').count().is(gt(0)),
    sack(shift).by(constant([1,0,0,0])).out(`left'),
    sack(shift).by(constant([1,0,0,0])).
     sack(reflect).by(constant(`lr'))),
   choose(out(`right').count().is(gt(0)),
    sack(shift).by(constant([0,1,0,0])).out(`right'),
    sack(shift).by(constant([0,1,0,0])).
     sack(reflect).by(constant(`lr'))),
   choose(out(`up').count().is(gt(0)),
    sack(shift).by(constant([0,0,1,0])).out(`up'),
    sack(shift).by(constant([0,0,1,0])).
     sack(reflect).by(constant(`ud'))),
   choose(out(`down').count().is(gt(0)),
    sack(shift).by(constant([0,0,0,1])).out(`down'),
    sack(shift).by(constant([0,0,0,1])).
     sack(reflect).by(constant(`ud'))))).
 times(26)
\end{verbatim}
\end{small}

\begin{figure}[h!]
	\centering
	\includegraphics[width=0.48\textwidth]{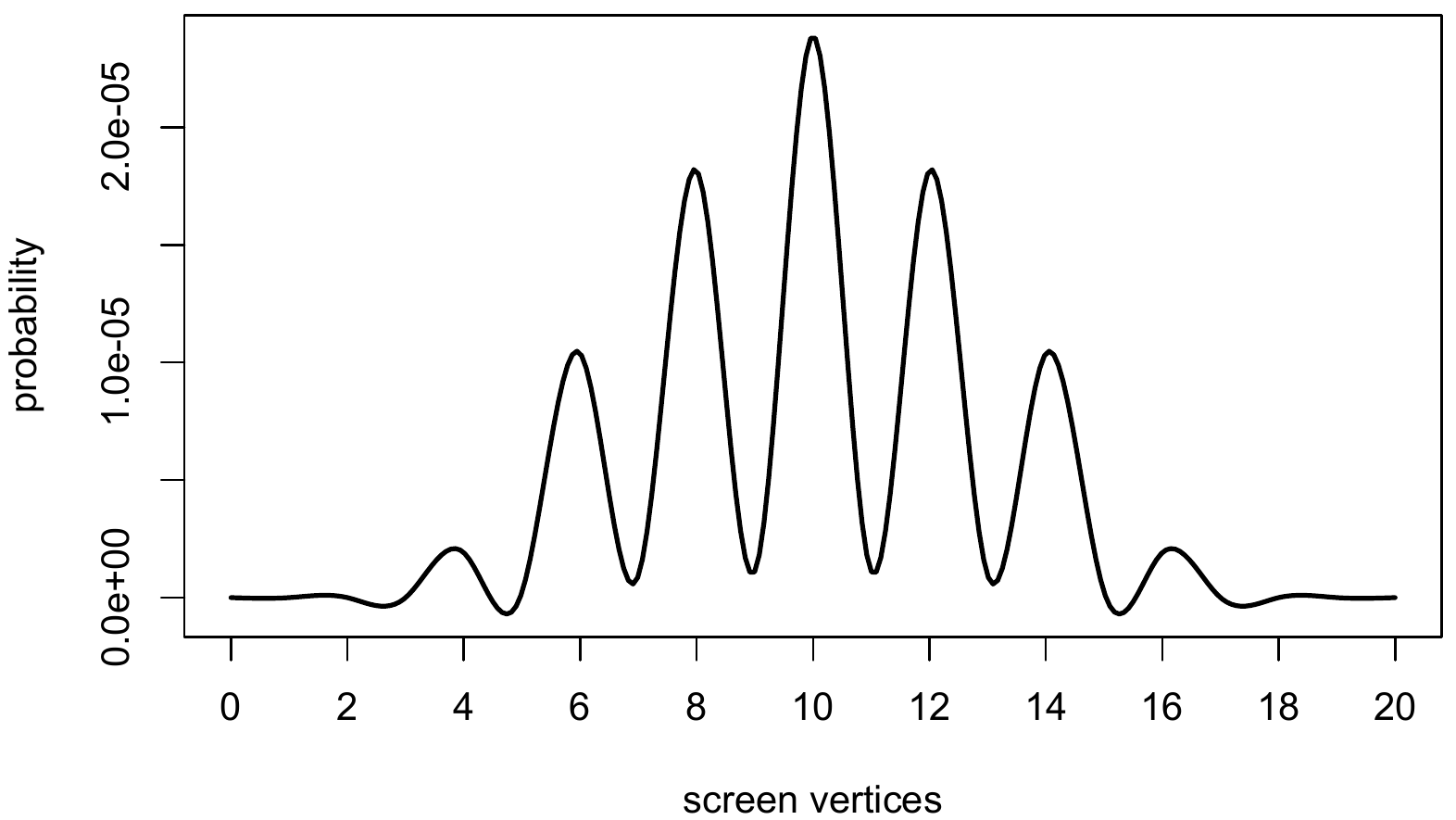}
	 \caption{\label{fig:double-slit-screen}The probability distribution on the top $20$ lattice vertices after 24 iterations. The wavefunction experiences quantum interference and, as realized in nature, a ``sliced" probability distribution is yielded.}
\end{figure}

Figure \ref{fig:double-slit-screen} shows the probability distribution on the top $20$ lattice vertices (the light sensitive film) after $26$ iterations. As expected, the waves emanating from the two slits interfere with each other yielding a constructive peak at the center vertex with diminishing intensity towards the ends.

\begin{figure}[h!]
	\centering
	\includegraphics[width=0.40\textwidth]{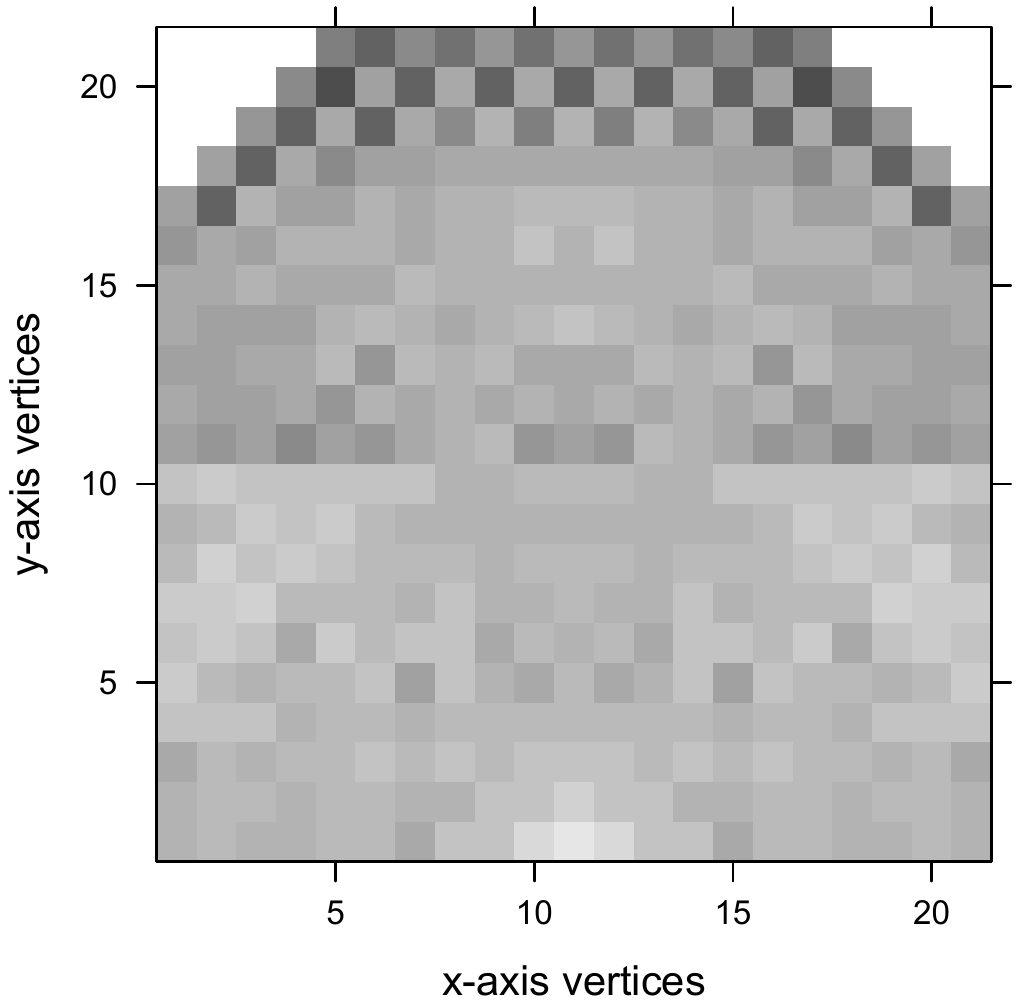}
	 \caption{\label{fig:double-slit-grid}The total probability distribution of the traverser's location over the $20 \times 20$ vertex lattice at iteration 26.}
\end{figure}

There are 2 types of boundary conditions in the double slit screen experiment. The sides of the lattice and the mid-screen barrier. However, because of the \texttt{reflect} function, the barriers reverberate the wavefunction back while still preserving the unitary nature of the quantum process. Figure \ref{fig:double-slit-grid} shows the probability distribution at iteration 26, where lighter gray vertices (squares) have a higher probability than the darker gray vertices. The white vertices on the top left and right have no probability and denote the wave front boundary.

Finally, it is important to emphasize that this experiment's results vary depending on the coin used, the size of the lattice, the depth and width of the slits, as well as the number of iterations allowed until measurement \cite{doubleslit:oliverira2007}.

\section{Quantum Semantics in Classical Traversals}

All quantum walk algorithms share the same common description.
\begin{enumerate}
  \item Put a single (classical) traverser into a position basis and spin basis.
  \item Alter the traverser's spin state using a unitary operation which conserves the total spin of the system.
  \item Given the new spin state, create (quantum) traversers at the respective adjacent vertices with shifted spin.
  \item If two (quantum) traversers meet at the same vertex, sum their spin states using pairwise vector addition and make one (quantum) traverser. 
  \item Repeat steps 2 through 4 for each (quantum) traverser for some number of iterations.
  \item Generate a probability distribution based on the sum of the absolute squares of all (quantum) traversers across the graph.
  \item Sample that probability distribution and yield a single (classical) traverser in a position and spin basis.
\end{enumerate}
The distinction between a ``classical traverser" and a ``quantum traverser" is simply that the classical traverser is the initial and end state of the system when $|T| = 1$ and it is in a basis state (i.e.~at a particular vertex with a spin completely in one direction).\footnote{A nice mental distinction to make is that when $|T| = 1$  there exists a ``particle" (classical) and when $|T| > 1$, there exists a ``wave" (quantum).} This is a superficial distinction and as such, there is nothing fundamentally different about the two types of traversers in Gremlin. Furthermore, the requirement for wavefunction collapse is simply a requirement of the natural world and not something inherent to simulated quantum walks. By removing probabilistic sampling (steps 6 and 7), a quantum traversal is equivalent to a classical Gremlin traversal save for the notion of ``spin." The concept of spin can be leveraged in Gremlin (irrespective of ``quantum processing."). When spin is used to encode branch frequency (and not branch amplitude), then the quantum concepts presented thus far are useful in ``classical" Gremlin traversals.

\subsection{Intersection and Symmetric Difference with Traverser Spin}

The branches of a traversal determine the number of components in a traverser's spin. On a line graph, there are two options that are \texttt{union}-d together -- \texttt{out(`left')} and \texttt{out(`right')}. Thus, a two dimensional spin array is required. The first number in the array is the amount of spin-left and the second, the amount of spin-right. For the lattice example, a four dimensional array was required. Each dimension represented the amount of left-, right-, up-, and down-spin in the traverser. In general, the number of spin dimensions required is equal to the number of options or degrees of freedom in the traversal.

\begin{figure}[h!]
	\centering
	\includegraphics[width=0.26\textwidth]{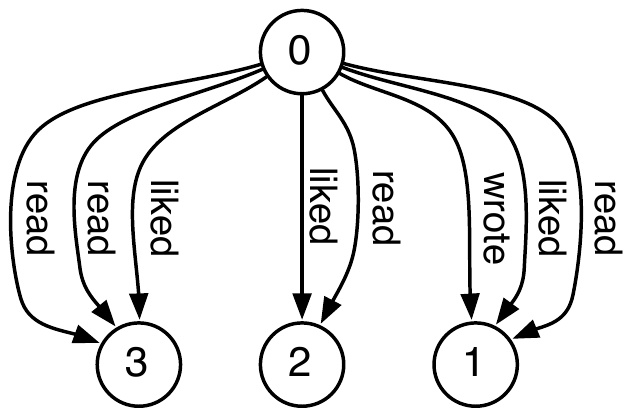}
	 \caption{\label{fig:simple-graph}A simple 4-vertex/8-edge property graph.}
\end{figure}

When the spin of a traverser is understood as its superposition in $\Psi$, then the topology of the graph and the traversal can be decoupled. In all the quantum experiments presented thus far, the traversal topology mirrored the graph topology. If the vertices had an outgoing \texttt{left}-edge, then the traversal had a respective \texttt{out(`left')}-step. In real-world property graphs, the graph structure is complex with different traversals identifying different features of the graph. For instance, the graph diagrammed in Figure \ref{fig:simple-graph} can contain \texttt{read}-, \texttt{wrote}-, and \texttt{liked}-edges throughout, but if a particular traversal is trying to identify the most central person in the implicit co-authorship graph, then the traversal will only contain \texttt{out(`wrote')} and \texttt{in(`wrote')} branches. Thus, the traverser is constrained to a particular subgraph of $G$ and that constraint is expressed in the branches of its traversal.

Next, unitary evolution is necessary in the natural world as the total spin of the system must be conserved. In the world of graph computing, this constraint is not required and in fact, if unitary evolution is abandoned, then traverser spin ``bookkeeping" denotes how many traversers were generated by each branch of the traversal. With this information it is possible to implement intersection and symmetric difference using traverser spin. To do this, three methods are defined. The methods \texttt{merge} and \texttt{shift} accomplish pair-wise vector addition and multiplication of an arbitrary dimension, respectively. The method \texttt{split} is analogous to the coins defined for quantum walks, but instead of conserving spin (unitary), it sums the vector and places that sum in every dimension. The understanding here is that if the traverser is in $[1,0,0]^\top$ spin prior to a three-branch traversal, then when it goes into spin superposition and thus, down all three branches, the traverser's spin/bulk is $[1,1,1]^\top$.
\begin{small}
\begin{verbatim}
merge = {a,b -> x = []; 
 (0..a.size()-1).each{ i -> x << a[i] + b[i] }; x }
shift = {a,b -> x = []; 
 (0..a.size()-1).each{ i -> x << a[i] * b[i] }; x }
split = { a,b -> x = []; 
 a.each{ x << a.sum() }; x }
\end{verbatim}
\end{small}
In the following traversal, a traverser is placed at vertex 0 of the graph diagrammed in Figure \ref{fig:simple-graph}. That traverser's spin is then \texttt{split} to be $[1,1]^\top$. Semantically, this means that the traverser has gone into a $\psi$-superposition as it will take both branches of the \texttt{union}-step. Each branch will \texttt{shift} the spin of the traverser to zero out the component of the opposing branch (i.e.~a projection). The \texttt{barrier}-step aggregates all the traversers and merges those traversers at the same location (as defined by the equivalence class $[t]$) into a single traverser via pairwise vector/array addition. Finally, the \texttt{map()}-step serves as a view into the state of each traverser where \texttt{t.get()} is the vertex location of the traverser and \texttt{t.sack()} is its respective sack (i.e.~spin).
\begin{small}
\begin{verbatim}
g.withSack([1,0],merge).V(0).
 sack(split).
 union(
  sack(shift).by(constant([1,0])).out(`read'),
  sack(shift).by(constant([0,1])).out(`wrote')).
 barrier().map{t -> [t.get(),t.sack()]}
\end{verbatim}
\end{small}
The output of this traversal over the toy graph in Figure \ref{fig:simple-graph} is provided below.
\begin{small}
\begin{verbatim}
==>[v[1], [1, 1]]
==>[v[2], [1, 0]]
==>[v[3], [2, 0]]
\end{verbatim}
\end{small} 
The spin state of the merged traverser provides enough information to determine intersection (i.e.~and) and symmetric difference (i.e~exclusive or), where
\begin{equation*}
\texttt{intersect}(t) = 
\begin{cases}
\texttt{true} &: \forall i \in \varsigma(t) \; i > 0 \\
\texttt{false} &: \text{otherwise}
\end{cases}
\end{equation*} 
and
\begin{equation*}
\texttt{symDiff}(t) = 
\begin{cases}
\texttt{true} &: \exists i \in \varsigma(t) \; i = \sum_j \varsigma(t)_j \\
\texttt{false} &: \text{otherwise}.
\end{cases}
\end{equation*} 
These predicates are expressed using a \texttt{filter}-step.\footnote{In this context, the Gremlin \texttt{filter}-step acts in a similar fashion as a \textit{polarized screen} to light. A polarized screen only allows light through that is at a particular spin.} For intersection,
\begin{small}
\begin{verbatim}
g.withSack([1,0],merge).V(0).
 sack(split).
 union(
  sack(shift).by(constant([1,0])).out(`read'),
  sack(shift).by(constant([0,1])).out(`wrote')).
 barrier().filter(not(sack().unfold().is(eq(0)))).
 map{t -> [t.get(),t.sack()]}
\end{verbatim}
\end{small}
The \texttt{filter} only allows those traversers to pass that don't have a sack component equal to 0. Given that only vertex 1 can be reached by both \texttt{read}- and \texttt{wrote}-edges, then the intersection of these two branches filters out vertex 2 and vertex 3.
\begin{small}
\begin{verbatim}
==>[v[1], [1, 1]]
\end{verbatim}
\end{small}

For symmetric difference, the \texttt{filter} is different as it must ensure that one and only one sack component is greater than 0.
\begin{small}
\begin{verbatim}
filter(sack().unfold().is(gt(0)).count().is(eq(1))).
\end{verbatim}
\end{small}
The result of the symmetric difference is below.
\begin{small}
\begin{verbatim}
==>[v[2], [1, 0]]
==>[v[3], [2, 0]]
\end{verbatim}
\end{small}

Note that the number of traversers generated by each branch is contained in the respective spin component. For instance, there are two \texttt{read}-edges from vertex 0 to vertex 3 and zero \texttt{wrote}-edges from vertex 0. Thus, the traverser located at vertex 3 has a spin state of $[2,0]^\top$. Given that the number of traversers generated by each branch is split amongst the spin component, a ``measurement" of the spin state of the traverser at its respective vertex location will collapse the spin to a frequency-based basis state. 
\begin{small}
\begin{verbatim}
norm = { a,b -> x = []; 
 x[0] = a.sum(); (0..a.size()-2).each{x << 0}; x }
\end{verbatim}
\end{small}
\begin{small}
\begin{verbatim}
g.withSack([1,0],merge).V(0).
 sack(split).
 union(
  sack(shift).by(constant([1,0])).out(`read'),
  sack(shift).by(constant([0,1])).out(`wrote')).
 barrier().
 filter(not(sack().unfold().is(eq(0)))).
 sack(norm).map{t -> [t.get(),t.sack()]}
\end{verbatim}
\end{small}
The result of the above intersection-traversal is below. The number of traversers after branching all fold to the first component/branch as the traversal is now back to a linear (non-branching) form.
\begin{small}
\begin{verbatim}
==>[v[1], [2, 0]]
\end{verbatim}
\end{small}

The above example had two branches (``read" and ``wrote") and thus, a length 2 spin vector. In the traversal below, there are three branches and thus, a length 3 spin vector is used.
\begin{small}
\begin{verbatim}
g.withSack([1,0,0],merge).V(0).
 sack(split).
 union(
  sack(shift).by(constant([1,0,0])).out(`read'),
  sack(shift).by(constant([0,1,0])).out(`wrote'),
  sack(shift).by(constant([0,0,1])).out(`liked')).
 barrier().map{t -> [t.get(),t.sack()]}
\end{verbatim}
\end{small}
The result of the traversal above is below. Using the appropriate \texttt{filter}, intersection or symmetric difference can be effected.
\begin{small}
\begin{verbatim}
==>[v[1], [1, 1, 1]]
==>[v[2], [1, 0, 1]]
==>[v[3], [2, 0, 1]] 
\end{verbatim}
\end{small}

The significance of this model is that there is no side-effect data structure that aggregates all the traversers in each branch and then does an intersection/difference of those aggregations. The traversal is fully functional as the branch statistics are encoded on the traverser's spin and thus, the ``aggregation" is distributed across the traversers and therefore, across the traversal flow.

\subsection{Distributing Global Data Structures Across Local Traverser Spin}

The ability to abandon side-effect ``bookkeeping" data structures is important because it enables steps to be purely functional (stateless). When steps are stateless, then they can more easily be executed in both a threaded and machine distributed manner. This is perhaps made more salient in the following example where an explicit side-effect can be abandoned in favor of a traverser spin representation.

A common pattern in graph traversing is to determine if a vertex has already been touched by a previous step in the traversal. For instance: ``\textit{Who are my friends' friends that are not my friends?}" To answer this question, it is important to know all of the person's friends, then, for each of those friends, determine their friends while excluding those friends-of-a-friend that are not the original person's friend. This is currently accomplished in Gremlin by using the \texttt{aggregate}-step which generates a side-effect data structure. In the example, this data structure is referenced by the variable $x$.
\begin{small}
\begin{verbatim}
g.V(0).out(`knows').aggregate(`x').
 out(`knows').where(not(within(`x')))
\end{verbatim}
\end{small}
The \texttt{aggregate(`x')}-step is a barrier in that it blocks until all traversers prior to it have passed through it. These traverser's locations (i.e.~friends) are stored in the set $x$. Once all the traverser locations have been aggregated, the barrier is ``drained" one traverser at a time by the \texttt{out(`knows')}-step which computes the friends-of-a-friend.\footnote{A barrier-step, like \texttt{aggregate} and \texttt{barrier}, ensures a breadth-first execution up to that point in the traversal. Gremlin's architecture supports traversals that go from breadth- to depth-first over the course of the computation.} The \texttt{where}-step filters out all those traversers that are at the same vertex location as any of those stored in $x$. This is not a purely functional operation. The problem is that $x$ is a ``global blackboard" that is accessed by all traversers. This limits what can be done in a distributed environment as the $x$ data structure is external to the traversal flow. However, the previous spin-based technique can be used to implicitly store $x$ in the traverser flow. The above non-functional traversal can be rewritten in a purely functional way. The final filter ensures that no merged traverser went down the first ``identity" branch (i.e.~$x$). 
\begin{small}
\begin{verbatim}
g.withSack([1,0],merge).V(0).out(`knows').
 sack(split).
 union(
  sack(shift).by(constant([1,0])),
  sack(shift).by(constant([0,1])).out(`knows')).
 barrier().
 filter(sack().unfold().range(0,1).is(eq(0)))
\end{verbatim}
\end{small}

The interesting aspect of this traversal is that there are in fact two branches to union. However, the first branch stays on the friend vertex and simply shifts the traverser's spin while the other branch shifts the spin and then moves to the friends-of-a-friend via the \texttt{out(`knows')}-step. While not necessary, the first branch could have ended with an \texttt{identity}-step to make this ``stall" more apparent.

\subsection{Applying Quantum Concepts to Classical Gremlin}

The Gremlin compiler makes use of \textit{traversal strategies} to introspect a traversal prior to its execution. The purpose of these strategies is to rewrite particular sequences of steps into a more optimal representation. For instance, \texttt{out().count().is(eq(5))} is rewritten to \texttt{out().limit(6).count().is(eq(5))} by the \texttt{RangeByIsCountStrategy}. It is possible to use such strategies to dynamically introduce spin into the traverser's sack definition. This would enable the user to simply write
\begin{small}
\begin{verbatim}
g.V(0).intersect(out(`read'),out(`wrote'))
\end{verbatim}
\end{small}
and have the compiled version use the aforementioned spin representation. The user would not have to be concerned with sack splitting, shifting, and filtering. In fact, the concept of spin is alien to the user and only serves as a functional optimization. Next, suppose the following abstract traversal, where $X$ and $Y$ are standard functional step sequences.
\begin{small}
\begin{verbatim}
g.X.intersect(out(`read'),out(`wrote')).Y
\end{verbatim}
\end{small}
In this situation, Gremlin would use classical constructs in the $X$-sequence, quantum constructs in the \texttt{intersect}-sequence, and then ``collapse" the system back to a single spin for the $Y$-sequence. In many ways, this is how quantum computers and classical computers are expected to interact. An algorithm will have classical and quantum components. The classical aspects compute up to the quantum part at which point, the classical data is sent to the ``quantum chip" for the quantum part of the algorithm to execute. When the quantum component is complete, the wavefunction is collapsed and the resultant basis state is fed back into the classical part of the algorithm. Gremlin can follow this model, save that a compiler strategy (not a user) would be responsible for breaking up the traversal into classical and quantum components for their execution using ``bulks" (classical) or ``spins" (quantum), respectively.

\section{Conclusion}

Gremlin is a graph traversal machine and language that can evaluate traversals on graphs represented on a single computer or across an arbitrarily large compute cluster. It can be used to execute any known algorithm and thus, is Turing Complete. Furthermore, Gremlin maintains a step library (instruction set) encompassing numerous graph traversal primitives. The expressivity of its step library enables it to conveniently represent and execute quantum walks. In order to do this, Gremlin traverser ``sacks" are endowed with a complex vector. These sacks undergo unitary evolution and in the process, yield constructive and destructive interference as the unitary operation ``rotates" them and merges them at vertex locations in the graph. Collapsing the wavefunction to a probability distribution and ultimate classical basis state is also conveniently expressed in Gremlin. The concept of quantum ``spin" has been demonstrated to be useful outside of pure quantum simulation, where unitary amplitude evolution is abandoned for a frequentist evolution. The use of spin allows typical non-functional Gremlin constructs to be represented in a purely functional way. Future work in this area will explore other aspects of Gremlin that can take advantage of quantum concepts as well as make it easier for quantum computing practitioners to leverage Gremlin for quantum simulation and potentially, as a general purpose quantum programming language. 

\begin{acknowledgements}
The authors would like to thank the Apache Software Foundation for their commitment to open source software and specifically for their support of the Apache TinkerPop project. TinkerPop has been developed and maintained by various individuals over the years and each individual's unique efforts have helped shape and grow Gremlin. Finally, the authors are grateful to Lynn Bender for the creation of the GraphDay conference and his support of the graph computing space.
\end{acknowledgements}

\bibliographystyle{abbrvnat}
\bibliography{../marko}

\end{document}